# A Low-Power, Low-Latency, Dual-Channel Serializer ASIC for Detector Front-End Readout

**Le Xiao[a,b], Datao Gong[b,*], Tiankuan Liu[b], Jinghong Chen[c], Qingjun Fan[c], Yulang Feng[c], Di Guo[b], Huiqin He[d], Suen Hou[e], Guangming Huang[a], Xiaoting Li[g,a,b], Chonghan Liu[b], Quan Sun[b], Xiangming Sun[a], Ping-Kun Teng[e], Jian Wang[b,f], Annie C. Xiang[b], Dongxu Yang[b,f], Jingbo Ye[b]**

[a] *Department of Physics, Central China Normal University,*
*Wuhan, Hubei 430079, P.R. China*

[b] *Department of Physics, Southern Methodist University,*
*Dallas, TX 75275, USA*

[c] *Department of Electrical and Computer Engineering, University of Houston,*
*Houston, TX 77004, USA*

[d] *Shenzhen Polytechnic,*
*Shenzhen 518055, China*

[e] *Institute of Physics, Academia Sinica,*
*Nangang 11529, Taipei, Taiwan*

[f] *State Key Laboratory of Particle Detection and Electronics, University of Science and Technology of China,*
*Hefei Anhui 230026, China*

[g] *Institute of High Energy Physics, Chinese Academy of Sciences, Beijing 10004, China*
*E-mail*: dgong@mail.smu.edu

ABSTRACT: In this paper, we present a dual-channel serializer ASIC, LOCx2, and its pin-compatible backup, LOCx2-130, for detector front-end readout. LOCx2 is fabricated in a 0.25-μm Silicon-on-Sapphire CMOS process and each channel operates at 5.12 Gbps, while LOCx2-130 is fabricated in a 130-nm bulk CMOS process and each channel operates at 4.8 Gbps. The power consumption and the transmission latency are 900 mW and 27 ns for LOCx2 and the corresponding simulation result of LOCx2-130 are 386 mW and 38 ns, respectively.

[*] Corresponding author.

## Contents

## 1. Introduction

When LHC upgrades to high luminosity, the trigger readout electronics of ATLAS Liquid Argon (LAr) Calorimeter will have to be redesigned. One of the challenges for Liquid Argon calorimeter trigger electronics upgrade is the need of transferring huge amounts of data from the detector to the control room within limited latency and power budgets [1]. This calls for a low-power, low-latency, high-speed serializer. For a serializer that carries 16-channels ADC data, the latency budget is 50 ns and the power budget is 1 W. The GBTX application-specific integrated circuit (ASIC) [2] is a potential candidate, but one of the reasons it cannot be used is that its latency is beyond the latency budget.

Based on a commercial 0.25-µm Silicon-on-Sapphire (SoS) CMOS process [3], we designed a dual-channel high-speed serializer ASIC named LOCx2 [4]. The SoS process was chosen because of its immunity to single event effects. The first fully functional prototype of LOCx2 performs well in normal operating conditions, but it has a duty cycle distortion (DCD) problem after irradiation. After analyzing potential sources that cause DCD, we revised the design and submitted a new prototype. Test results indicate that the problem has been solved. The problem found in the first prototype, the subsequent design improvement, and the test results of the new design are presented in this paper.

Besides the new prototype of LOCx2, we migrated the design to a 130-nm bulk CMOS process and named it as LOCx2-130. LOCx2-130 uses a silicon-proven modified analog core of GBTX with improved latency [5]. Simulation indicates that the latency of LOCx2-130 is comparable to that of LOCx2, but the power consumption of LOCx2-130 is less than half of that of LOCx2. The design and simulation results of LOCx2-130 are also presented in this paper.

The paper is organized as follows. In Section 2, we report the design and the test results of LOCx2. In Section 3, we present the design and simulation results of LOCx2-130. In Section 4 we summarize the paper.

## 2. LOCx2

LOCx2 is a dual-channel serializer and each channel operates at the data rate of 5.12 Gbps. Figure 1. is the block diagram of LOCx2. LOCx2 is composed of two encoders, two 16:1 serializers, two Current-Mode Logic (CML) drivers, a shared Phase-Locked-Loop (PLL), and an Inter-Integrated Circuit ($I^2C$) slave. LOCx2 supports two types of ADCs, an ASIC called Nevis ADCs [6] and a Commercial-Off-The-Shelf (COTS) device (part number ADS5272 [7] produced by Texas Instruments). Only the interface with Nevis ADCs is shown in the figure. The encoder, named as LOCic [8], encodes the ADC data for the serializer (16:1 multiplexer shown in the figure). The serialized data are sent to an optical module named MTx [9]. The PLL provides a 2.56-GHz clock signal to each serializer. The $I^2C$ slave is implemented for user configuration.

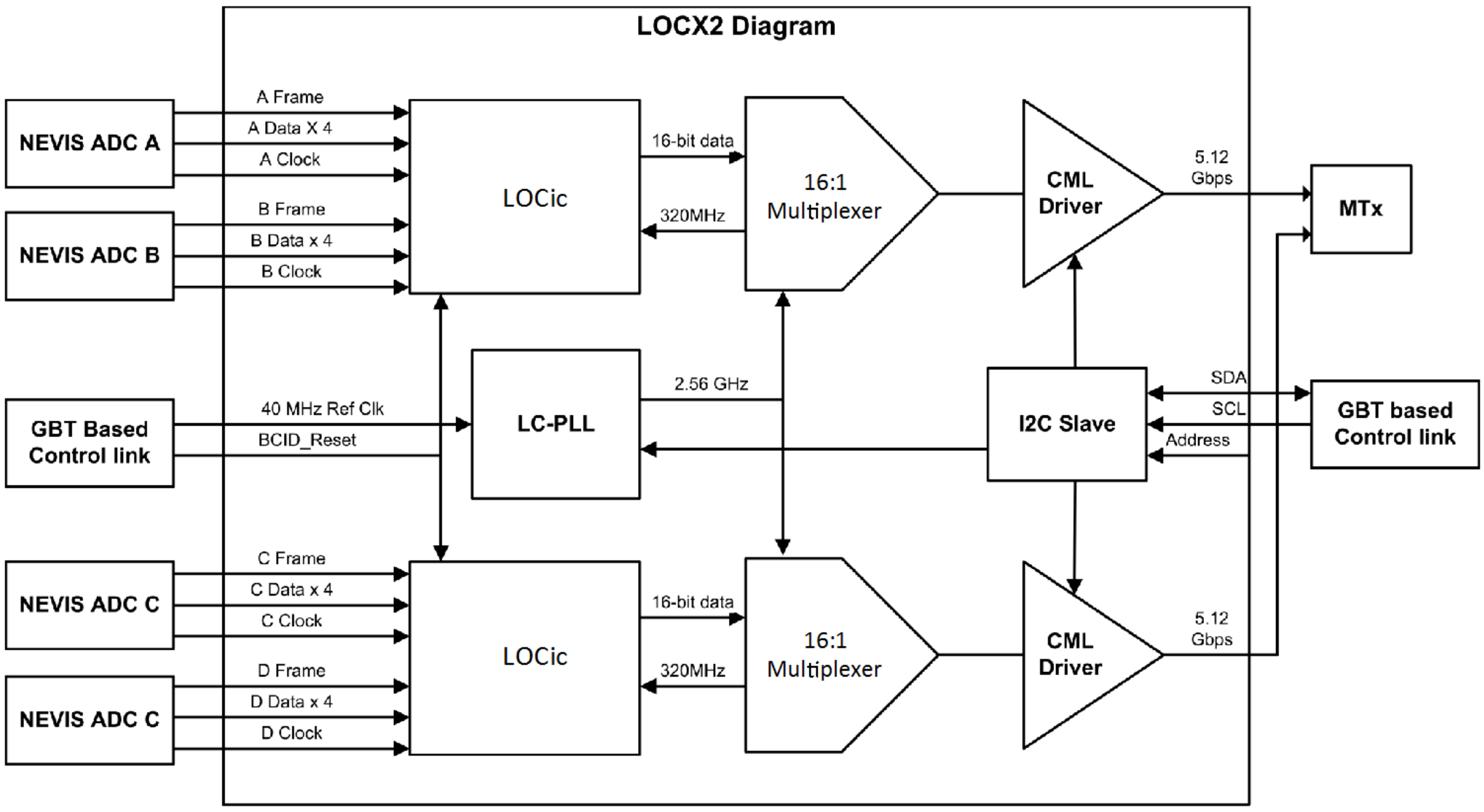

**Figure 1**. Block diagram of LOCx2.

### 2.1 Problem found in the previous prototype

The first fully functional prototype has been fabricated and tested. The prototype performs well in normal operating conditions, but suffers from a DCD problem after irradiation. Figure 2 shows the eye diagrams of two-channel output serial data before and after 20 krad irradiation in the X-ray, much less than the total dose requirement of the front-end electronic system in ATLAS LAr. In the figure, the cyan and green curves are eye diagrams of two serial output data and the purple curve is the test clock output of the PLL. The eye diagrams were observed by using a 40-MHz clock instead of a 5.12-GHz clock to trigger the oscilloscope in order to observe the DCD effect.

After analyzing all potential sources, we conclude that the DCD of the output serial data comes from the DCD of the 2.56-GHz clock used in the last stage of multiplexers. The schematic of the last stage of multiplexers is shown in Figure 3. If the duty cycle of the 2.56-GHz clock deviates from 50%, the output serial data will carry the DCD. This DCD transfer mechanics is shown in Figure 4.

The DCD on the clock comes from the mismatches of process parameters, such as threshold voltage and transistor size, resistive and capacitive loads. Furthermore, it is known that the transistor parameters change in radiation [3]. Radiation may strengthen the mismatch effects and deteriorate the duty cycle distortion. Because the foundry does not provide any mismatch model, we cannot perform precise simulation. However, by manually adding parameter mismatch, we have simulated the mismatch effects and our simulation results are qualitatively consistent with our observations. First, it is well known that the mismatch in a differential buffer causes a DC offset on the signal, resulting in DCD in the clock signal. Our simulation also indicates that the DCD of the differential clock buffers increases with process mismatches. For example, a mismatch of 2.5% in capacitive load of each clock buffer can result in a DCD of 7% in the output clock of the clock buffer chain. This is consistent with a mild DCD sporadically observed even before the irradiation test. Second, our simulation indicates that with the same amount of capacitive load mismatch, a threshold voltage mismatch of 0.2 V can induce an extra 50% increase in the DCD. Third, our simulation confirms an observed phenomenon that an increase of the power supply voltage can mitigate the DCD.

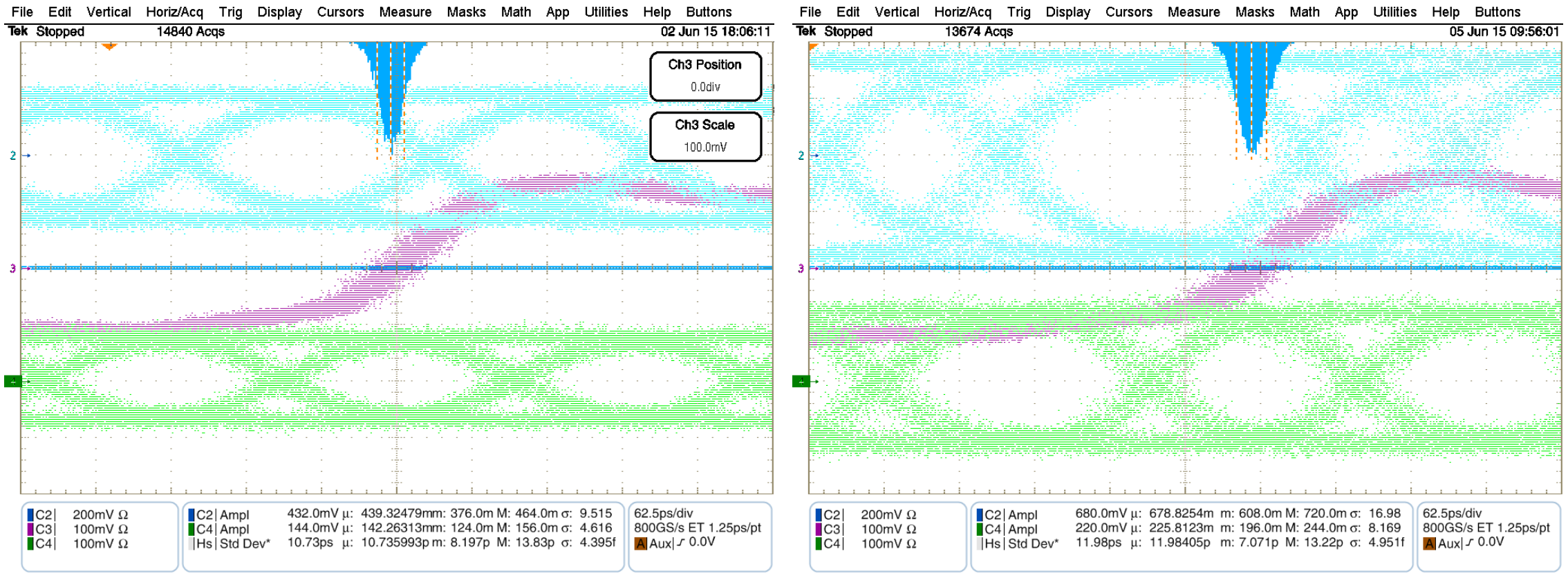

**Figure 2.** Eye diagrams before (left) and after (right) irradiation.

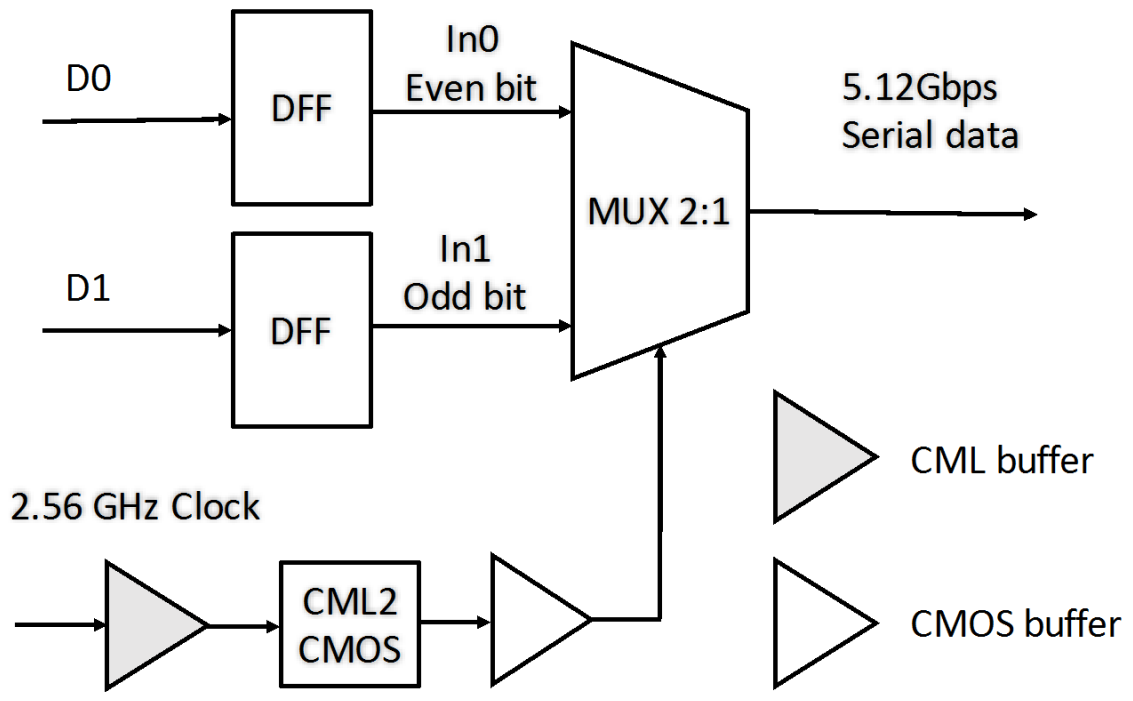

**Figure 3.** Structure of the last stage of multiplexers.

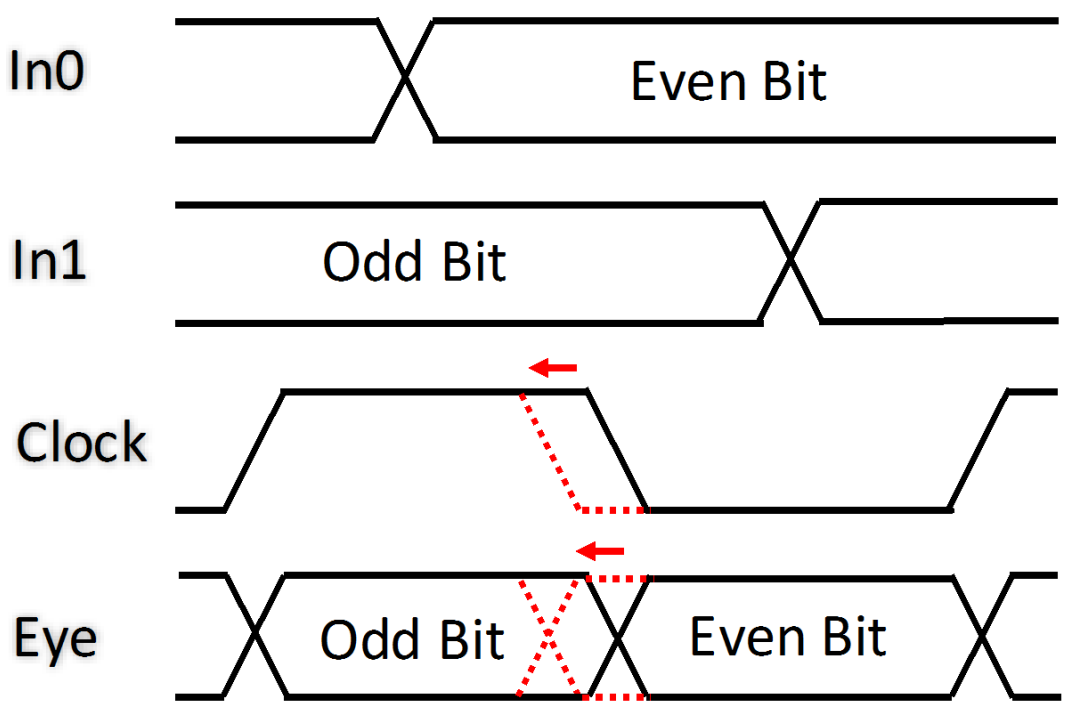

**Figure 4**. DCD transferring.

## 2.2 Design of the new prototype

Based on the analysis of the problem found in the previous prototype, we re-designed the clock distribution circuits to improve the DCD robustness of the CML clock buffers. Figure 5 is the schematic of the CML clock buffer with its biasing circuit. We used AC coupling between the Voltage Controlled Oscillator (VCO) of the LC-PLL and the first CML buffer to remove the potential DC offset from the VCO. The feedback biasing circuits keeps the common mode voltage of all CML clock buffers at the same level of 1.5 V (i.e., 0.6×VDD shown in the schematic), set by two resistors.

In order to mitigate the mismatch effects in the CMOS clock buffers and the mismatch effects unrelated to the DC offset of the CML buffers, we also added duty cycle correction (DCC) and clock alignment circuits [10]. These circuits recover the duty cycle of the CMOS clock signal to 50% if DCD occurs. The DCC circuit and clock aligners are shown in Figure 6. The inverter with a resistive feedback adjusts the common mode level of the CMOS clock signal to an optimal duty cycle. Simulation results indicate that two stages of such DCC circuits in series can correct the duty cycle from 70% or 30% to a few percent from 50%. A clock aligner is used to align the phase of the two complementary CMOS clock signals. Note that the driving strength of the CMOS inverters between the two complementary signals is half of that of the normal CMOS inverters. Simulation results indicate that two stages of clock aligners can decrease the phase delay of the two complementary signals to 20% of that before the clock aligners.

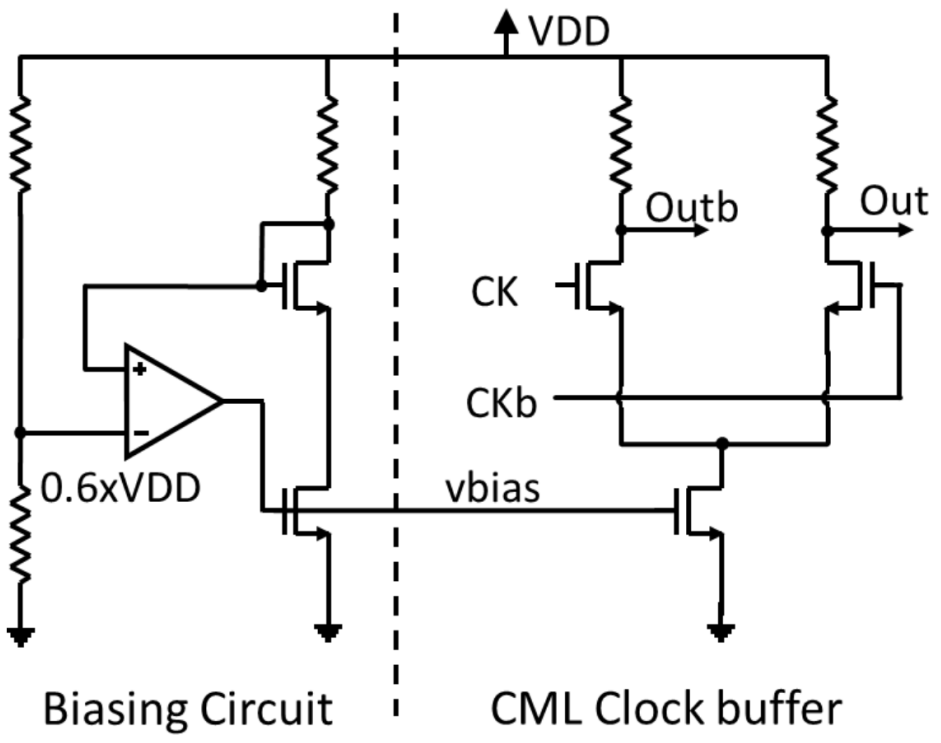

**Figure 5.** Schematic of a CML buffer with biasing.

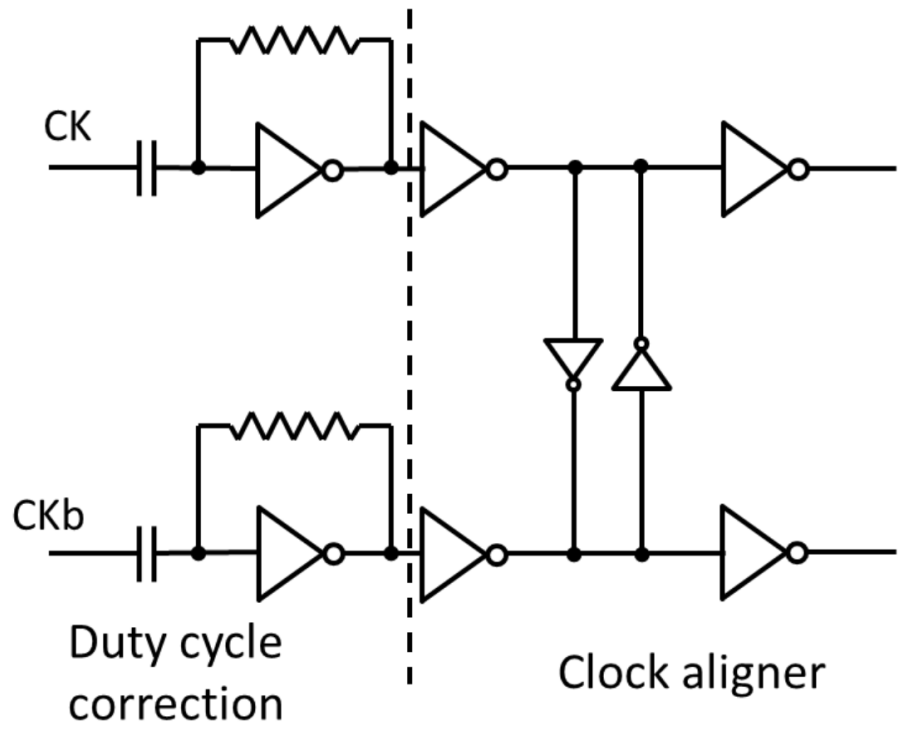

**Figure 6**. DCC and clock alignment.

### 2.3 Test results of the new prototype

The revised design has been fabricated and tested. The chip functions as expected. We did not observe any obvious DCD before or after irradiation under the dose required for the LAr Calorimeter. Figure 7 shows the eye diagrams after irradiation. Note that the eye diagram was observed with a QFN packaged chip in a socket installed on the board, which limits the bandwidth. The eye diagram with a QFN packaged chip soldered directly on the board is shown in Figure 8. We are planning a Single-Event-Effect (SEE) test before the end of this year.

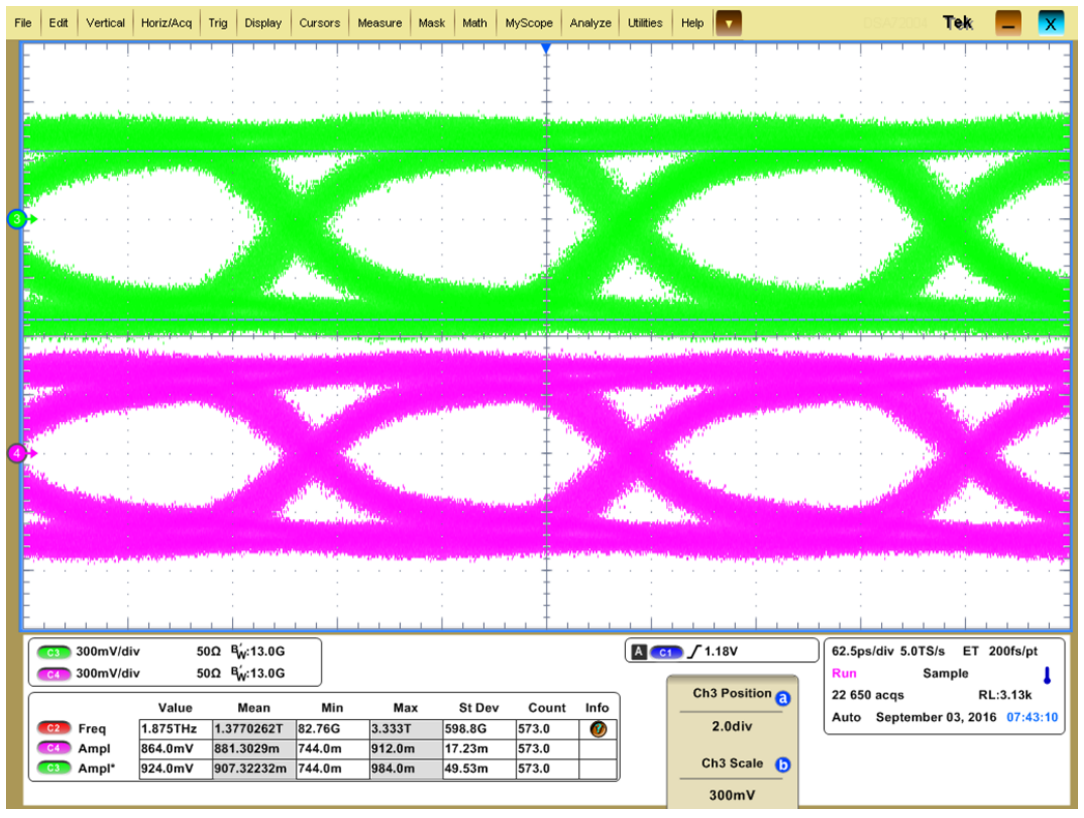

**Figure 7.** Eye diagram test with a QFN packaged chip in a socket installed on PCB after irradiation.

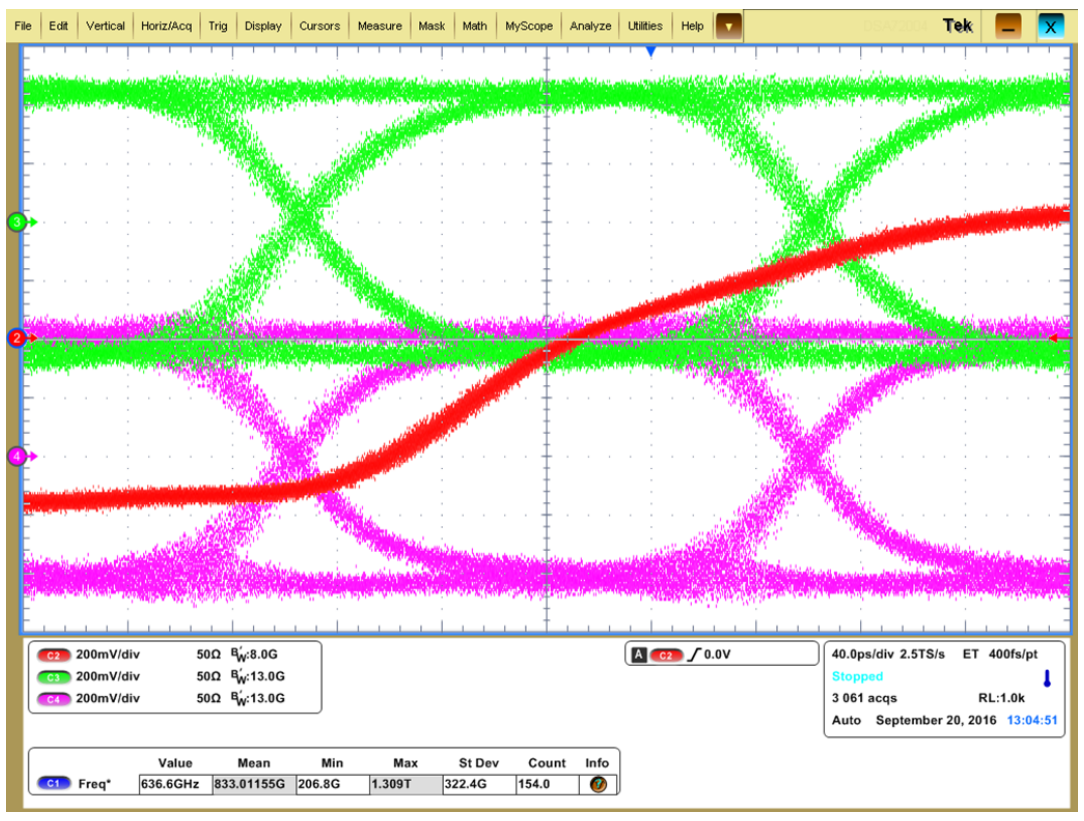

**Figure 8**. Eye diagram test with a QFN packaged chip soldered on PCB directly.

## 3. Design of LOCx2-130

We designed a pin-compatible backup, LOCx2-130, based on a 130-nm bulk CMOS process. LOCx2-130 is a dual-channel serializer. Each channel operates at 4.8 Gbps. LOCx2-130 uses a single power supply of 1.5 V, while LOCx2 uses 2.5 V.

### 3.1 Design of LOCx2-130

LOCx2-130 is designed to reuse the silicon proven components and the test facilities as much as possible. We utilized the GBTX analog core, which has been modified to a 30:1 serializer at 4.8 Gbps, named as TDS [5]. The TDS analog core includes a PLL and a high-speed 30:1 serializer. The TDS analog core has shorter latency than the original 120:1 serializer. We modified the analog core so that two serializers share a single PLL. This modification significantly reduces the power consumption.

The data frame defined in LOCx2 was verified in the previous LOCx2 prototype. For LOCx2-130, we modified the frame definition due to the data rate downgrade from 5.12 Gbps to 4.8 Gbps. Figure 9 shows the frame definition of LOCx2 and LOCx2-130. LOCx2-13 uses the same 8-bit frame header, including 4-bit fixed code “0101” and 4-bit Pseudo-Random Binary Sequence (PRBS) code as LOCx2. LOCx2-130 supports both the Nevis ADC and ADS5272, the same as LOCx2 does. Both ADCs have a resolution of 12 bits, but the Nevis ADC uses 14 output bits for calibration. LOCx2-130 has two operating modes, the data mode and the calibration mode. In the data mode, LOCx2-130 uses 96-bit payload for 8-channel ADCs and 16-bit Cyclic Redundancy Check (CRC) code to protect the payload. In the calibration mode, the user can turn off the CRC protection and carry 14-bit ADC data.

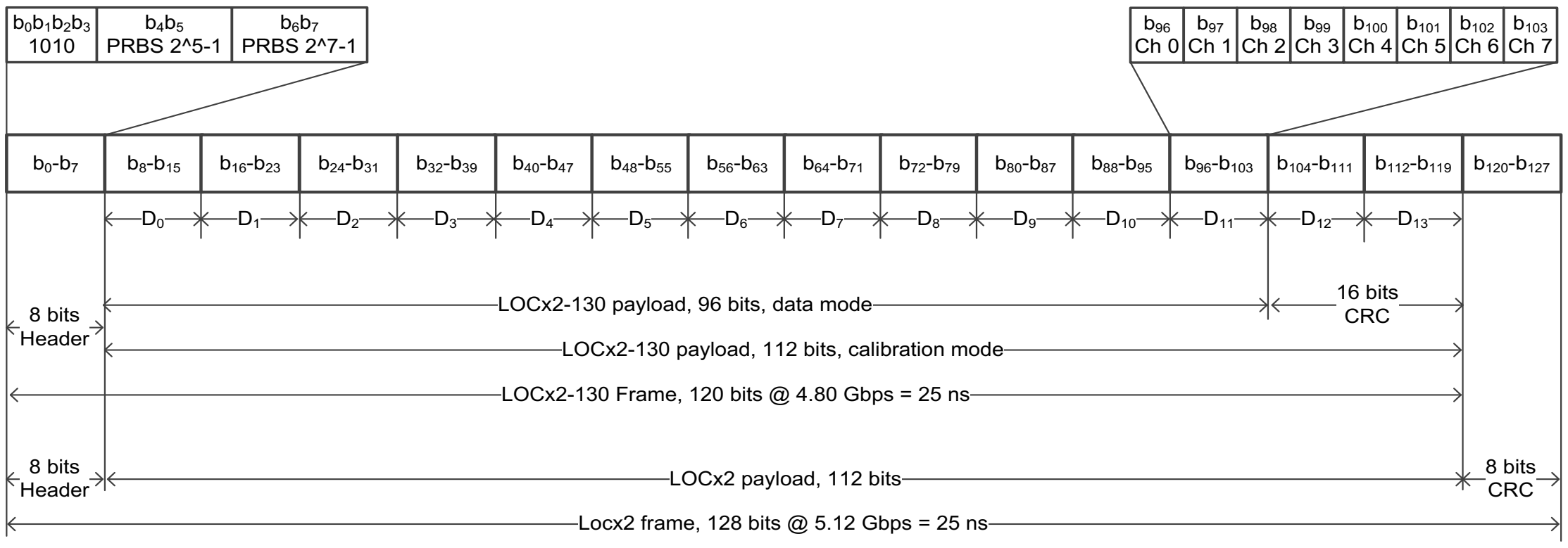

**Figure 9**. Frame definition of LOCx2 and LOCx2-130.

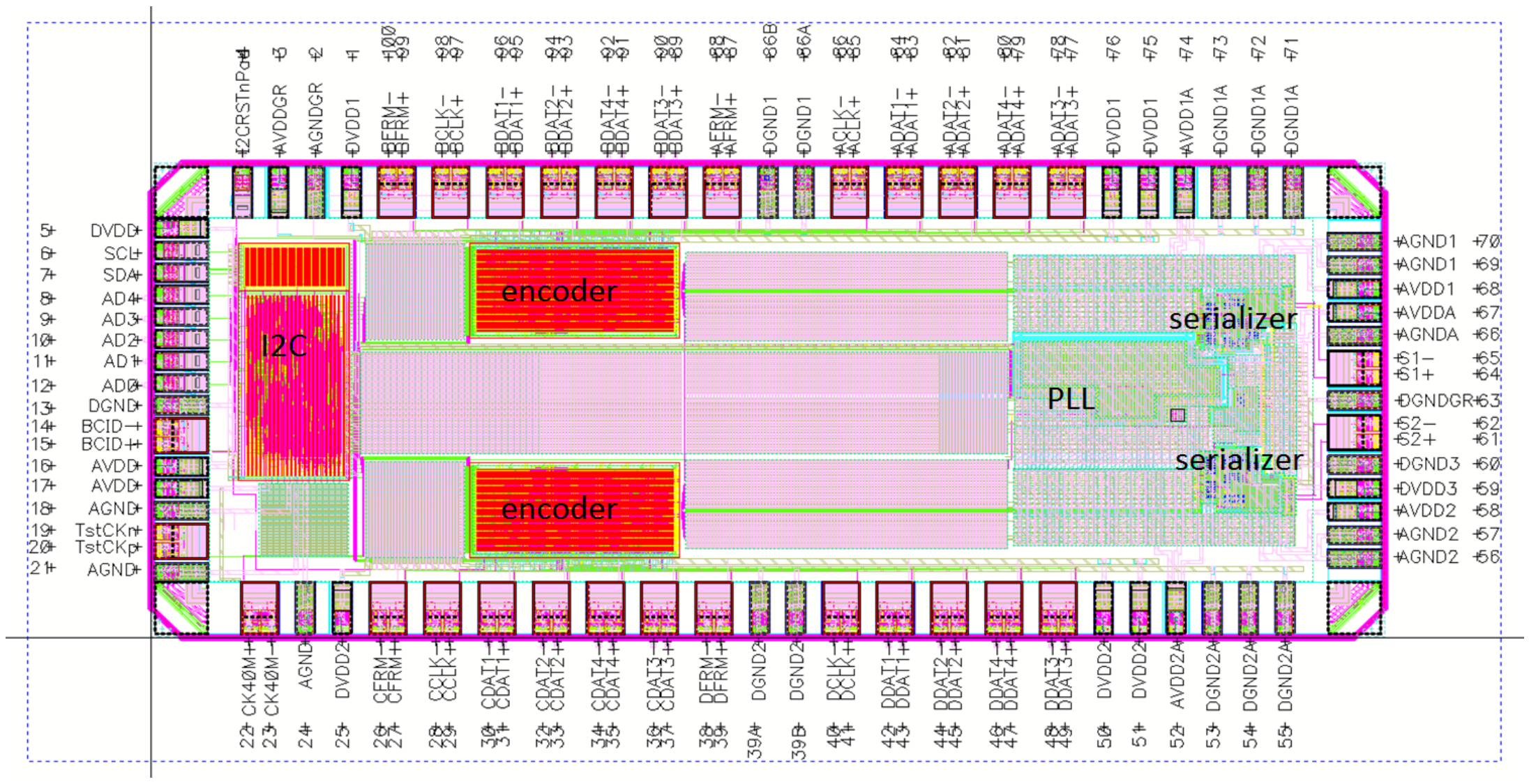

**Figure 10**. Layout of LOCx2-130.

Figure 10 is the layout of LOCx2-130. The analog core (the PLL and the two 30:1 serializers) is located at the right of the floor plan and has the largest decoupling capacitors (pink area in the figure) than other circuits. The digital functional blocks, including the Scalable Low Voltage Signaling (SLVS) receivers (located on the top and bottom edges) for the input signals, two encoders, and the $I^2C$ slave, are noisier than the analog core. We carefully isolated the substrates of these circuits and provide separated power supply and ground for each of them.

## 3.2 Simulation results of LOCx2-130

LOCx2-130 has been submitted for fabrication. Simulation indicates that the power consumption of LOCx2-130 is about 386 mW, less than half of LOCx2. The latency of LOCx2-130 is comparable to that of LOCx2 and within the latency specification. The comparison of LOCx2-130 and LOCx2 is shown in Table 1.

**Table 1.** Comparison of LOCx2 and LOCx2-130.

| | LOCx2 | LOCx2-130 |
|---|---|---|
| Process | 0.25-µm SoS | 130-nm CMOS |
| Number of serializer channels | 2 | |
| Data rate per channel (Gbps) | 5.12 | 4.8 |
| Latency (ns) | 24.1 - 27.3 | 34.4 - 40.7 |
| Power consumption (mW) | 950 (measured) | 386 (simulated) |
| Supply voltage (V) | 2.5 | 1.5 |
| Supported ADCs | Nevis ADC and ADS5272 | |
| Slow control interface | $I^2C$ | |
| Die size ($mm^2$) | 3.860 × 6.036 | 2.000 × 5.000 |
| Package | 100-pin QFN | |

## 4. Conclusion

We have designed and tested a dual-channel serializer ASIC, LOCx2 in a SoS process. The test results indicate that the improved design solved the DCD problem found in the previous prototype. We have designed a backup and pin-to-pin compatible prototype, LOCx2-130, in a 130-nm CMOS process. Simulation results show that the power consumption of LOCx2-130 is less than half of that of LOCx2, while the latency of LOCx2-130 is comparable to that of LOCx2.

## Acknowledgments

This work is supported by US-ATLAS R&D program for the upgrade of the LHC, the US Department of Energy Grant DE-FG02-04ER1299, and the National Natural Science Foundation of China under Grant No. 11375073 and 11220101005. We are grateful to Drs. Jinhong Wang and Junjie Zhu of University of Michigan for help on providing full TDS design and setup of the 130 nm CMOS process PDK, Drs. Hucheng Chen and Hao Xu of Brookhaven National Laboratory, Dr. Nicolas Dumont Dayot of LAPP, and Dr. Bernard Dinkespiler of CPPM for beneficial discussions in the encoder/decoder implementation and system integration.